\documentclass[10pt]{article}
\newcommand{\gtrsim}{\raise.3ex\hbox{$>$\kern-.75em\lower1ex\hbox{$\sim$}}\ }
\newcommand{\lesssim}{\raise.3ex\hbox{$<$\kern-.75em\lower1ex\hbox{$\sim$}}\ }
\usepackage[aps,prb]{revtex}
\usepackage{version2}
\usepackage{epsfig}
\begin{document}
\begin{title}
Statistical Analysis of X-ray Speckle at the NSLS
\end{title}
\author{Ophelia~K.~C.~Tsui and S.~G.~J.~Mochrie}
\begin{instit}
Center for Materials Science and Engineering,
Massachusetts Institute of Technology, Cambridge, MA 02139
\end{instit}
\author{L.~E.~Berman}
\begin{instit}
National Synchrotron Light Source, Brookhaven National Laboratory, Upton, NY 11973
\end{instit}
\begin{abstract}
We report a statistical analysis of the static speckle
produced by illuminating a disordered aerogel
sample by a nominally coherent x-ray beam at
wiggler beamline X25 at the
National Synchrotron Light Source.
The results of the analysis allow us to determine that
the coherence delivered to the X25 hutch is within
35\% of what is expected.
The rate of coherent photons is approximately two times
smaller than expected on the basis of the
X25 wiggler source brilliance.
\end{abstract}
\receipt{June 6, 1997.}
\onecolumn
\section{Introduction}

Insertion-device-based synchrotron x-ray sources
present us with the exciting opportunity to
carry out x-ray intensity fluctuation
spectroscopy (XIFS) measurements
to investigate the dynamics of condensed matter on molecular length scales
\cite{brauer:95,dierker:95,albrecht:96,mochrie:96}.
The success of these experiments, and others requiring coherent
x-rays, depends crucially on being able to employ
x-ray beams of the highest possible brilliance.
In principle, the number of x-ray photons per second in a
coherent beam is simply and
directly related to the source brilliance.
However, in planning XIFS experiments, it is imperative to quantify
whether the theoretical ideal is achieved at the sample under study.
If the brilliance/coherence at the sample under study
is less than expected or cannot be fully utilized,
steps may then be taken to remedy the situation.
We also note that the current interest in ``fourth generation'' synchrotron
sources derives in part from novel coherent x-ray experiments,
which may become possible with the extreme brilliance of these sources.
To make use of a fourth generation source, the promised
coherence must be delivered to the sample under study.

Recently, Abernathy and coworkers \cite{abernathy:96}
have introduced a method for conveniently quantifying coherence.
In brief,
a static, strongly-scattering aerogel sample is illuminated by a
coherent x-ray beam, prepared by means of a pinhole, immediately
upstream of the sample \cite{sutton:91,cai:94}.
The scattered x-rays give rise to
the speckle that is characteristic of
a disordered medium under partially-coherent illumination.
The statistical properties
of the observed speckle are
then analyzed to determine the coherence
properties of the illumination.
Perfect coherence corresponds to a single mode of the electromagnetic
field.
It turns out that it is possible to provide a good description of
the experimental statistics of the observed partially-coherent
x-ray speckle pattern
by supposing it to be the intensity sum
of several independent ``Gaussian '' modes of the
electromagnetic field, each of which contributes
its own independent perfectly-coherent speckle pattern.
The number of contributing modes ($M$) succinctly
specifies the coherence of the sample illumination and detection.
In this context, the variance of the intensity distribution
is $\beta = 1/M$, while the often-employed
second factorial moment is
$1 + \beta = 1 + 1/M$.
Incoherent illumination corresponds to $M =\infty$, {\em i.e.}
$\beta = 0$.
In the context of an XIFS measurement, $\beta$ is the expected
zero-time intercept of the normalized, baseline-subtracted
time autocorrelation function, while
the visibility of intensity fluctuations is $\sqrt{\beta}$.
We will use whichever of $\beta$ and $M$
seems the most convenient.

In the present paper, we follow this procedure to
determine the coherence of the beam delivered into
the hutch at wiggler beamline X25, at the
National Synchrotron Light Source (NSLS).
We find that the coherence at X25
is within 35\% of the theoretical ideal.
However, the rate of coherent photons is $\sim 2$ times
smaller than expected.
The origin of the discrepancy is uncertain, although
beryllium windows are candidates
for degrading the source brilliance \cite{snigirev:96}.
There were in addition graphite filters,
and multilayer monochromator crystals in the beamline.
However, the discrepancy here is much smaller than that
found in the earlier measurements performed at the European Synchrotron
Radiation Facility (ESRF) \cite{abernathy:96}. We
do not understand the difference between the NSLS and ESRF results.

The layout of the present paper is as follows.
In Section \ref{experiment}, we describe our experimental methods.
In Section \ref{results}, we present the measured
aerogel speckle and our analysis of its statistical properties.
We conclude in Section \ref{conclusion}.

\section{Experimental details}
\label{experiment}
The layout of beamline X25 as it pertains to our
experiment is as follows. The source is
a 27-pole wiggler, which is 1.6~m long.
There is
a set of pyrolytic graphite filters,
of total thickness 167~$\mu$m,
13~m downstream of the source.
The first beryllium window of thickness 254~$\mu$m.
(IF-1 ultra-high purity grade, but unpolished),
is just downstream of the graphite filters at 13~m.
7~keV x-rays within a bandwidth of 1.5\% were selected by
a vertically-diffracting, tungsten-boron carbide multilayer pair,
located at a distance of 18~m.
The multilayers are each known
to have a root-mean-square surface roughness of less than 5~${\rm \AA}$,
and a figure error of 2~arcseconds or less.
The final beryllium window of thickness 127 microns.
(IF-1 purity and buff-polished to a root-mean-square surface
roughness of 0.4~$\mu$m  or better on both surfaces)
was at a distance of about 26~m.
Adjustable slits were located immediately downstream of the second
beryllium window, and were set to 4.2~$\mu$m in the horizontal
and 8.9~$\mu$m in the vertical, as determined
by measurements of the full-width-at-half-maximum (FWHM)
of the Fraunhofer diffraction patterns in the horizontal
and vertical, respectively (data not shown).
Passing through this aperture were $\sim 5 \times 10^7$ nominally-coherent
photons per second.
The aerogel sample, 8~cm downstream of the slits,
was the same sample as employed by Abernathy {\em et al}
\cite{abernathy:96}, and was kindly supplied by Dr. Norbert Mulders
of the Pennsylvania State University.
The sample was contained in a evacuated cell with kapton windows for
x-ray access.
Between the sample and detector was an evacuated flight path of length 2.3~m,
also with kapton windows.
The detector -- a Princeton Instruments CCD camera, employing
direct x-ray detection --
was located 2.5~m downstream of the sample.
The CCD pixel size was $22 \times 22~\mu{\rm m^2}$.
The characteristics of this
detector have been analyzed and discussed in detail by Mainville {\em et al.}
\cite{mainville:96}.

\section{Results}
\label{results}
\subsection{Aerogel speckle}
Fig.\ \ref{fig:speckle} shows a rainbow-scale image of
the small angle scattering
obtained for a 95\%-void aerogel sample. These data are the sum of
twenty 100~s. exposures, obtained sequentially. The speckle pattern
found for each exposure was the same within counting statistics and
the time correlation function of the 20 exposures was constant. Therefore,
we believe that the incident beam was sufficiently stable
during the experiment.  The scattering
is comprised of a well-understood intensity envelope,
which is strongly modulated
by virtue of the partially-coherent illumination.
This is speckle.
The details of the speckle pattern
depend on the exact distribution of material in the aerogel.
The axes are labeled by CCD pixel number. The direct beam would
occur at pixel $(257,-27)$, in a region of the CCD that is masked
by a beam stop to prevent illumination of the CCD by the direct beam.
The wavevector range spanned by the illuminated region of the CCD
is $-0.0081$ to $0.0079~{\rm \AA^{-1}}$ in the horizontal
and 0.0021 to $0.0173~{\rm \AA^{-1}}$ in the vertical.

\begin{center}
\epsfig{figure=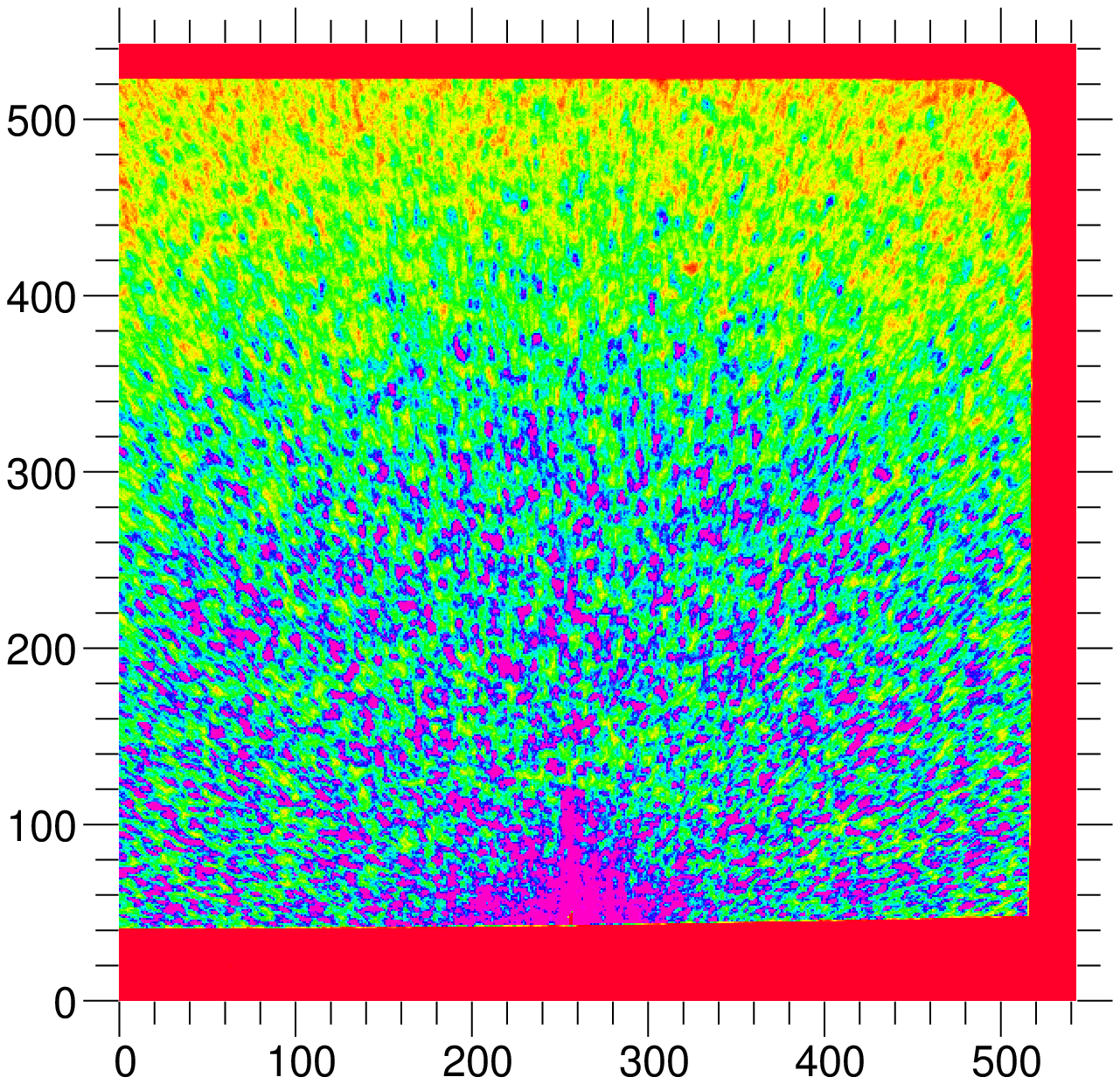,width=3.1in}
\end{center}
\figure{
Aerogel speckle pattern.
\label{fig:speckle}
}
\begin{center}
\epsfig{figure=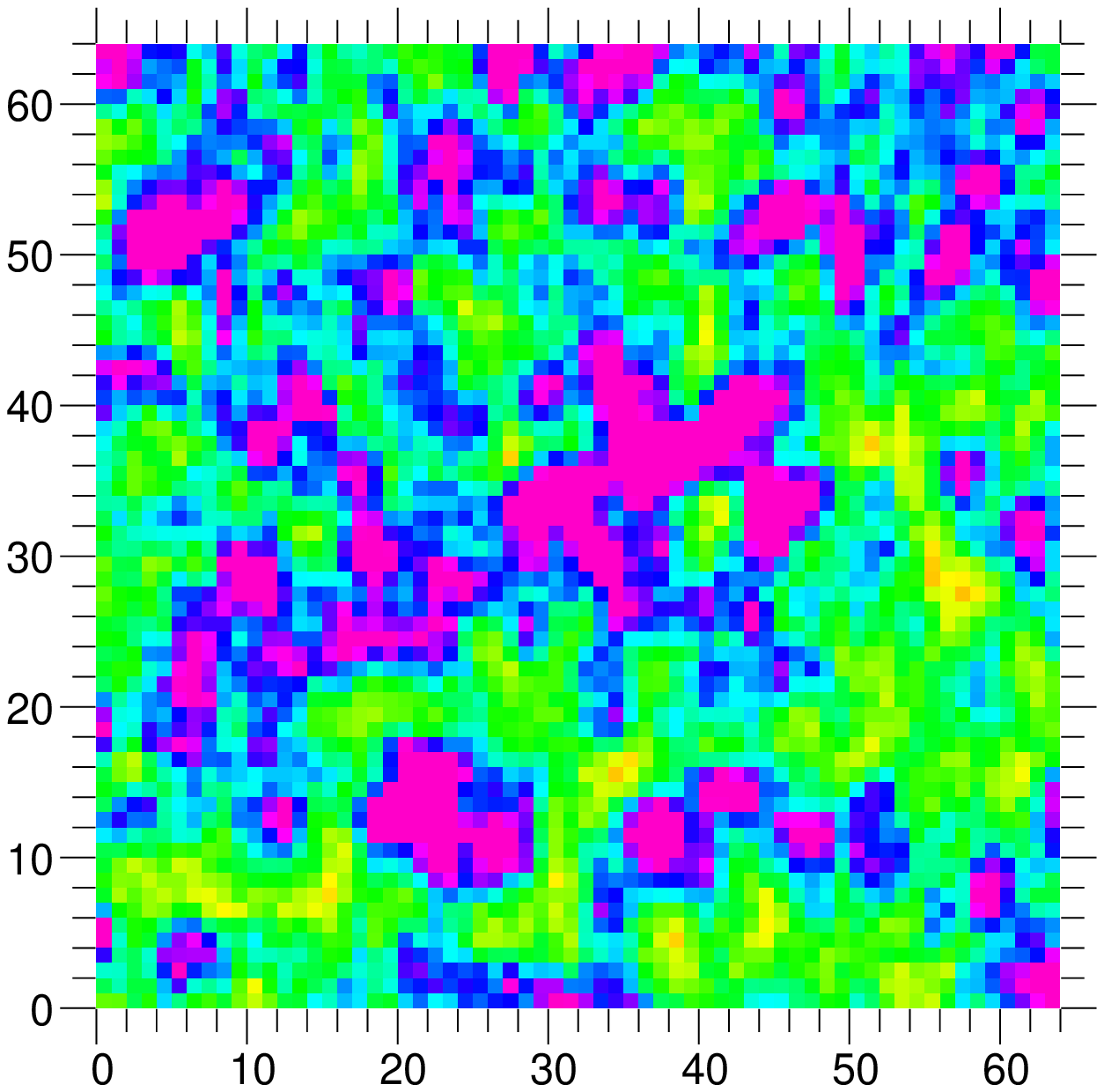,width=3.1in}
\end{center}
\figure{
Detail of the aerogel speckle pattern at 0.006~${\rm \AA^{-1}}$.
\label{fig:detail}
}
Fig.\ \ref{fig:detail} shows a $64 \times 64$ pixels sub-frame 
centered at a wavevector of $0.006~{\rm \AA^{-1}}$ in the
horizontal direction.
Inspection of Fig.\ \ref{fig:detail} reveals that the intensity does not
fluctuate from pixel to pixel, but rather it varies over a
scale of a few pixels.
We will quantify this observation below,
but we may already be confident that the spatial resolution of the detector
does not inadvertantly average the intensity fluctuations --
that is, the detection preserves the speckle visibility.
A striking feature of the speckle is 
a radial streaking, that becomes more
pronounced at larger scattering wavevectors.
This feature is due to the wavelength distribution of the incident
beam and will likewise be quantified below.
Fig.\ \ref{fig:azimuth} shows the azimuthally-averaged intensity for the
speckle shown in Fig.\ \ref{fig:speckle}. This represents the intensity
envelope.
\begin{center}
\epsfig{figure=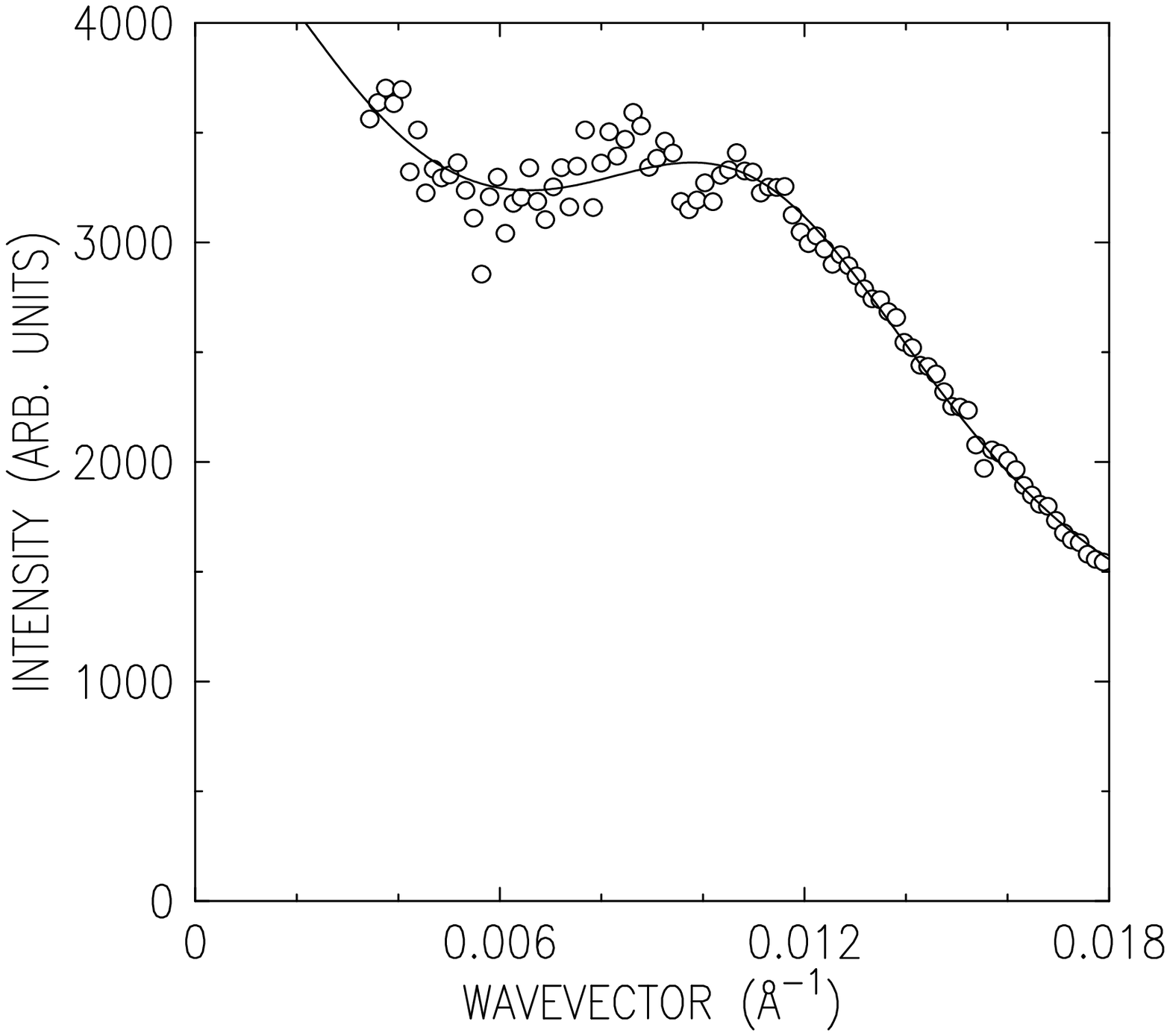,width=3.1in}
\end{center}
\figure{
Aerogel scattering intensity vs. wavevector
obtained by azimuthally
averaging the CCD image of Fig.\ \ref{fig:speckle} (open circles).
The line is a guide-to-the-eyes.
\label{fig:azimuth}
}

\subsection{Probability density of the speckle intensity}
\label{first-order}
The most characteristic feature of speckle is the large
point-to-point intensity variation, as evidenced
in Fig.\ \ref{fig:speckle}. Fig.\ \ref{fig:pI}(a)
and Fig.\ \ref{fig:pI}(b)
show the probability density of the speckle intensity
determined from sub-frames of $64 \times 64$ pixels
($0.002 \times 0.002~{\rm \AA^{-2}}$) centered at
wavevectors of $0.006~{\rm \AA^{-1}}$ and $0.016~{\rm \AA^{-1}}$,
respectively, obtained by histogramming the observed intensities
and normalizing by the respective mean intensities of the different
regions.
The insets show the intensity distributions on a logarithmic scale.
As might be anticipated, there is a wide distribution of
intensity -- far wider than would be expected for counting statistics.
Over the range of wavevectors included
in each sub-frame, the intensity of the scattering envelope
is approximately constant for wavevectors less that 0.01~${\rm \AA^{-1}}$.
Above 0.01~${\rm \AA^{-1}}$, the intensity changes with Q by a
noticeable amount.  However, within the wavevector
range of each sub-frame, the variation is no more than
$\sim 12$\% from its mean.  We found that for a sub-frame 
taken about Q$=0.016\AA^{-1}$, where it is the most susceptible
to this error,
the result is essentially unchanged by first normalizing the region
by the intensity envelope. Therefore, we conclude that 
throughout the wavevector
range we have studied, the intensity envelope makes
no contribution to the width of the intensity distribution.

The speckle produced by a coherent (single-mode) source
is expected to
display an exponential intensity distribution \cite{dainty:74}
{\em i.e.} $p_1(I)=(1/\bar{I})e^{-I/\bar{I}}$,
as a result of a Gaussian distribution of the electromagnetic
field strength about zero:
Because the field strength of the scattered wave
at a point in space consists of contributions
from each of the many electrons in the sample, the central limit
theorem of statistics implies
a Gaussian distribution of field strength.
Evidently, the experimental distributions in Fig.\ \ref{fig:pI}
do not conform to this form.
However, let us suppose that the observed partially-coherent
speckle may be considered to be the intensity
sum of $M$ independent coherent speckle
patterns, each with the same field strength. That is, $M$ 
modes of the electromagnetic
field contribute to the observed intensity.
It can be easily derived that 
the probability density of the
intensity in this case is  given by \cite{dainty:74}
\begin{equation}
p_M(I) =
\int \delta(I-\sum_{j=1}^{j=M}I_j) \prod_{j=1}^{j=M}p_1(I_j) dI_j
= M^M (I / \bar{I})^{M-1} e^{-M I / \bar{I}} / (\Gamma(M) \bar{I}).
\label{eq:pI}
\end{equation}
\begin{center}
\epsfig{figure=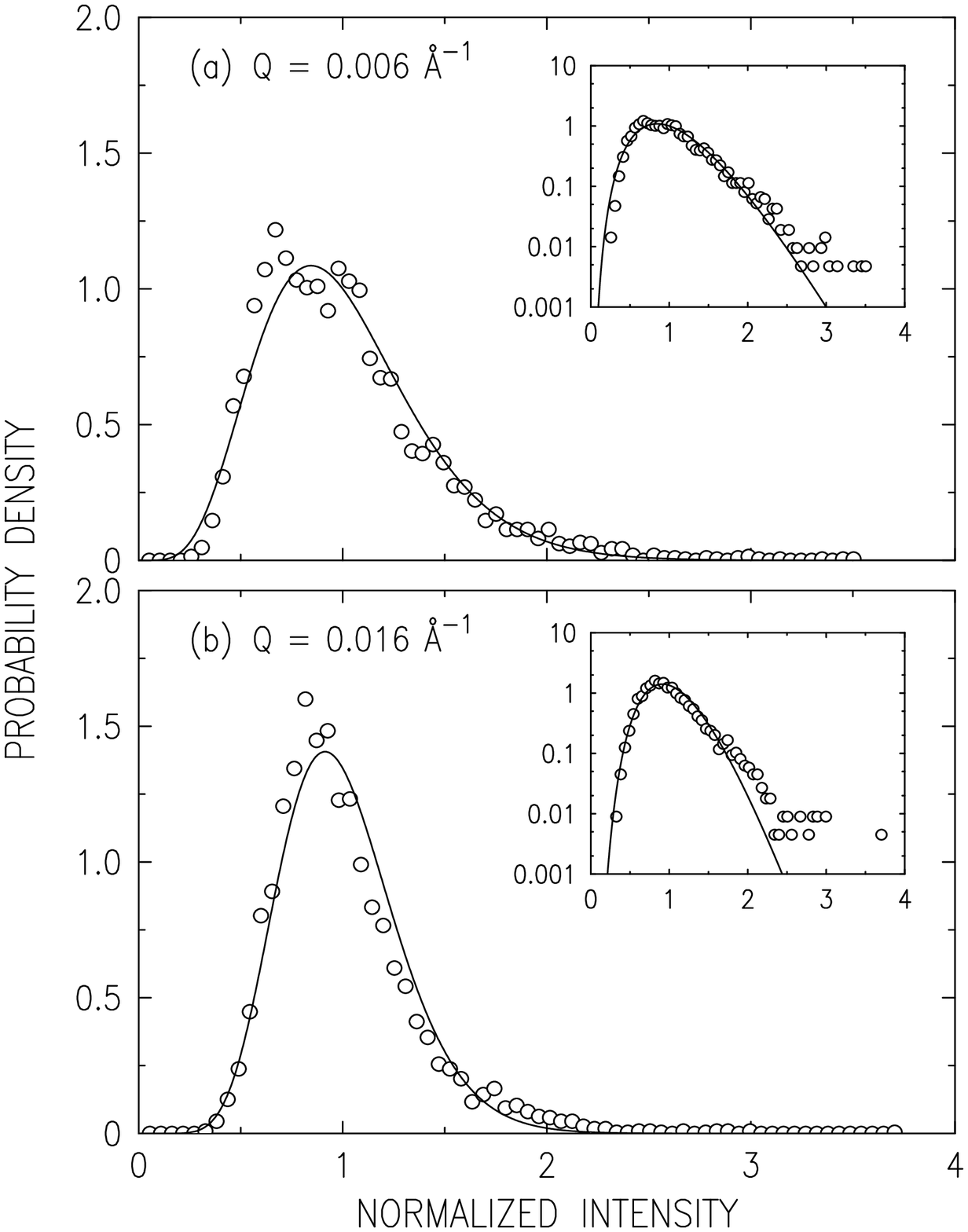,width=2.8in}
\end{center}
\figure{
Probability density of the speckle intensity
(a) 0.006~${\rm \AA^{-1}}$ and (b) 0.014~${\rm \AA^{-1}}$.
\label{fig:pI}
}
The distribution of intensity described
by Eq.\ \ref{eq:pI} has mean $\bar{I}$ and standard deviation
$\sigma_I = \bar{I} / \sqrt{M}$. This latter result implies that the
speckle visibility is $1/\sqrt{M} = \sqrt{\beta}$
for an intensity distribution described by Eq.\ \ref{eq:pI}.

Eq.\ \ref{eq:pI} is formally sensible for non-integer values of $M$.
The solid lines in Fig.\ \ref{fig:pI} show the best fit of
Eq.\ \ref{eq:pI} with variable $M$ to the experimental data.
The values of $M$ ($\beta$) so-obtained are $\sim 6$ ($\sim 0.17$)
and $\sim 11$ ($\sim 0.08$)
for $0.006~{\rm \AA^{-1}}$ and $0.016~{\rm \AA^{-1}}$, respectively.
An inspection of Fig.~\ref{fig:pI} reveals that the model peaks at 
an intensity that is slightly higher than the actual distribution does. 
Overall, however, the model form
provides a good description of the experimental distributions.
Fig.\ \ref{fig:MvsQ} shows the best fit values of $M$ plotted
versus wavevector for the distributions of Fig.\ \ref{fig:pI}
and for analogous distributions obtained at intermediate wavevectors.
\begin{center}
\epsfig{figure=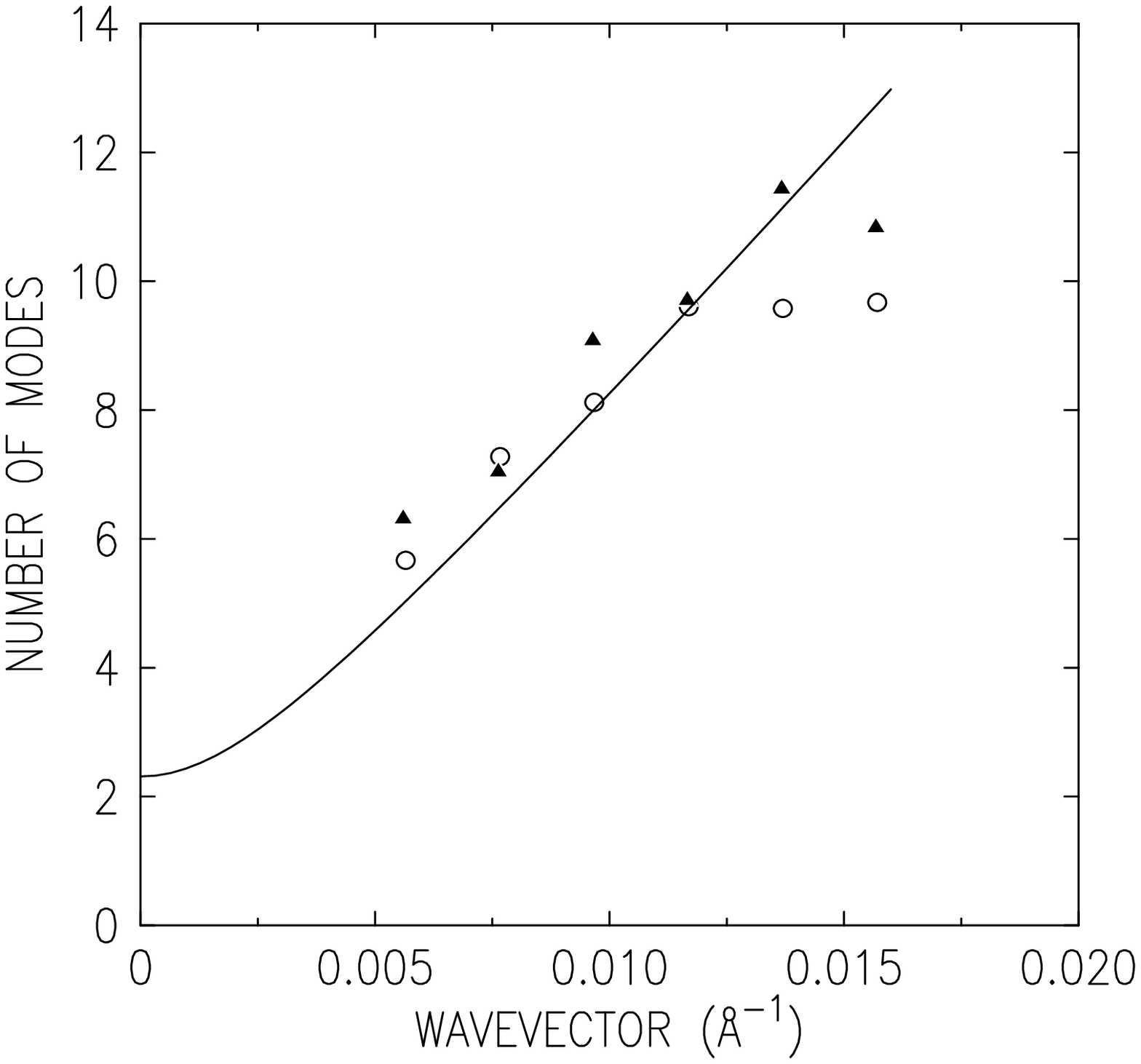,width=3.1in}
\end{center}
\figure{
Number ($M$) of coherent speckle patterns contributing
to the observed partially-coherent speckle pattern vs.\ wavevector.
Data were determined from fits to the intensity distribution with
variable $M$ (solid triangles), and from the speckle autocorrelation
(open circles).
\label{fig:MvsQ}
}


\subsection{Speckle intensity autocorrelation}
\label{second-order}
To quantify the speckle visibility independently of a model,
to quantify the speckle size,
and
to demonstrate that Poisson counting statistics do not
play a role in the observed intensity distributions,
we may calculate the
autocorrelation of the speckle between different pixels.
Fig.\ \ref{fig:autocorrelation}(a) and
Fig.\ \ref{fig:autocorrelation}(b)
show the normalized
autocorrelation versus pixel separation in the radial and tangential directions
for the sub-frames centered at $0.006~{\rm \AA^{-1}}$
and $0.014~{\rm \AA^{-1}}$, respectively.
It is worth noting that the
autocorrelation in the limit of zero pixel separation
is $1 + \sigma_I^2 / \bar{I}^2 = 1 + \beta$.
We believe that the autocorrelation's deviation from unity far from zero
reflects incomplete averaging as a result of the
limited size of the sub-frame considered ($64 \times 64$ pixels).
\begin{center}
\epsfig{figure=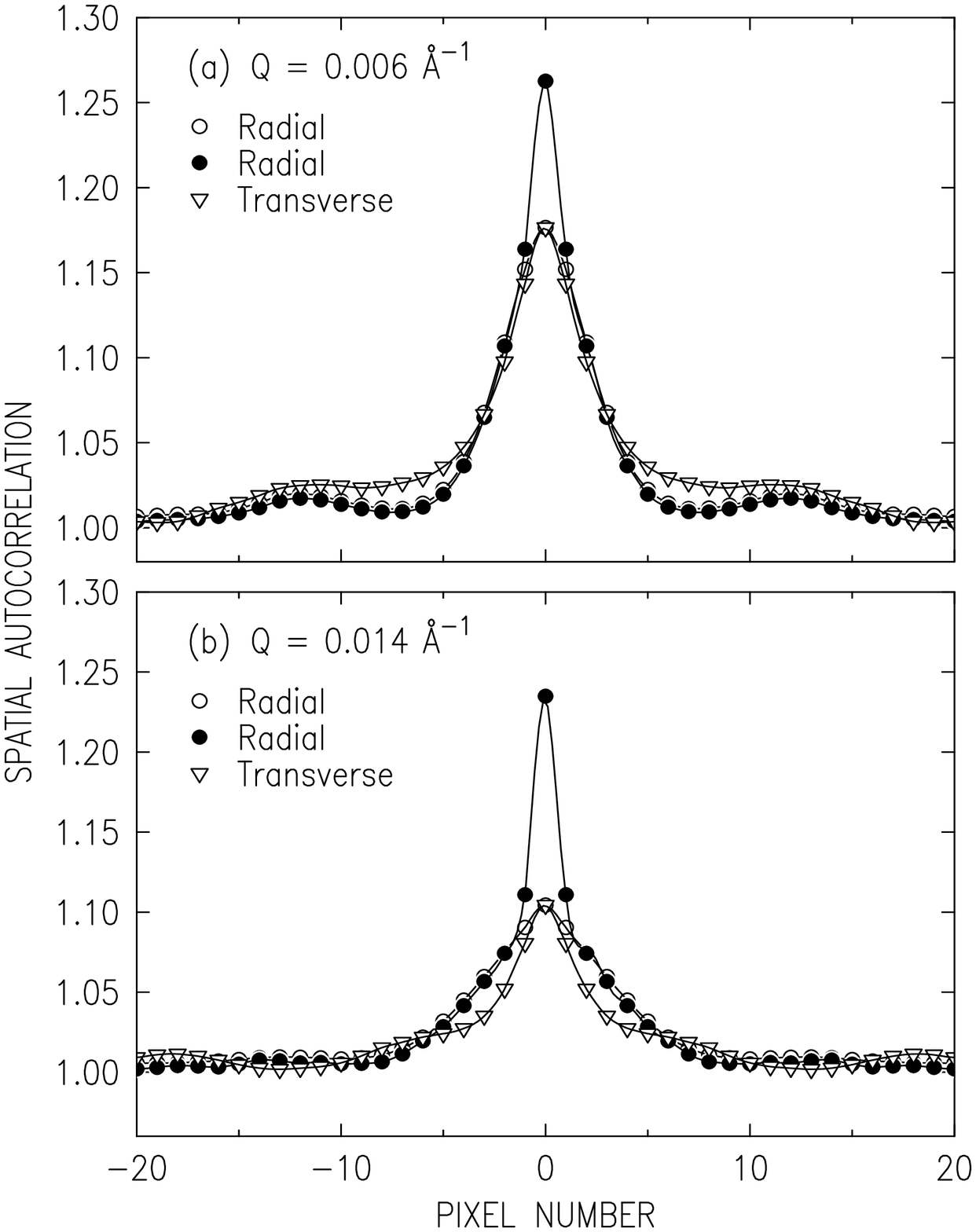,width=3.1in}
\end{center}
\figure{
Normalized
autocorrelation versus pixel separation in the radial (open and solid
circles) and tangential (open triangles) directions
near (a) 0.006~${\rm \AA^{-1}}$ and (b) 0.014~${\rm \AA^{-1}}$.
\label{fig:autocorrelation}
}

The normalized autocorrelation was calculated in two ways.
First,
the 20 exposures were summed together and the normalized autocorrelation
of the summed image was calculated. The results for the radial and
tangential correlations are shown as
open circles and open triangles, respectively.
Second, the normalized autocorrelation of each
exposure was calculated. The results of averaging the 20 individual
normalized autocorrelations in the radial direction
are shown as solid circles. Evidently, the autocorrelations
in the radial direction
obtained in these different ways overlap, except near zero.
The same is true for the tangential direction (data not shown
for the second method).

The difference between the autocorrelations obtained by the two
methods originates in the
20-times-larger contribution of Poisson counting statistics
to the intensity fluctuations for the second method.
Specifically,
since the Poisson contribution to
the variance is given by the number of photons counted \cite{mainville:96},
the difference between the two autocorrelations 
divided by the autocorrelation at large displacements
has an (integrated) amplitude equal to 19/20 times the inverse of the
average number of photons per pixel per 100 seconds counting time
\cite{dufresne:95}.
Thus, for example, near $0.006~{\rm \AA^{-1}}$, there were
approximately 6 x-ray photons per pixel per 100 seconds.
By examining the unnormalized autocorrelation, we may deduce
that there are approximately 630 detector units per 7~keV x-ray photon.
This result -- 90 detector units per keV -- agrees well
with the earlier, more detailed analysis of Ref.\
\onlinecite{mainville:96}, which deduced 94 detector units per keV
from measurements using 8.3~keV x-ray photons.

It is clear from Fig.\ \ref{fig:autocorrelation}
that Poisson counting statistics do not contribute
significantly to the autocorrelation of the summed image.
Therefore, the autocorrelation of the summed image
reflects accurately the speckle size and the speckle visibility.
On this basis, we may infer that counting statistics do not contribute
significantly to the intensity distributions of Fig.\ \ref{fig:pI}.
The open circles in Fig.\ \ref{fig:MvsQ} correspond to the
values of $M$ determined from $1/M = \sigma_I^2 / \bar{I}^2$
for the summed image autocorrelation at zero pixel separation. They
agree well with the values deduced from the
fits to the model intensity distributions (see Fig.~\ref{fig:MvsQ}).

It is important to note that the Poisson contribution to the
autocorrelation directly determines the detector resolution,
{\em i.e.} the amount of ``cross-talk'' between neighboring pixels.
For the autocorrelations near
$0.006~{\rm \AA^{-1}}$,
the Poisson contribution to the autocorrelation displays a FWHM
of 1 pixel, and the speckle contribution to the autocorrelation
displays a FWHM of 4 pixels.
It follows that the effective detector resolution is
indeed much smaller than the speckle size, and does not
significantly diminish the observed intensity fluctuations.
Since the speckle autocorrelation widths increase with wavevector,
this will also be true at larger wavevectors than 
$0.006~{\rm \AA^{-1}}$.

In Fig.~\ref{fig:autocorrelation}, it may be seen that the
tangential width is approximately the same at the two wavevectors
shown, but that the radial 
width is considerably larger at the
larger wavevector, reflecting the observed radial
streaking of the speckle, seen in  Fig.\ \ref{fig:speckle}.
This observation is quantified in Fig.\
\ref{fig:FWHM}, which shows the full-width-at-half-maximum (FWHM)
of the background-subtracted normalized autocorrelation
in the radial and tangential directions.
\begin{center}
\epsfig{figure=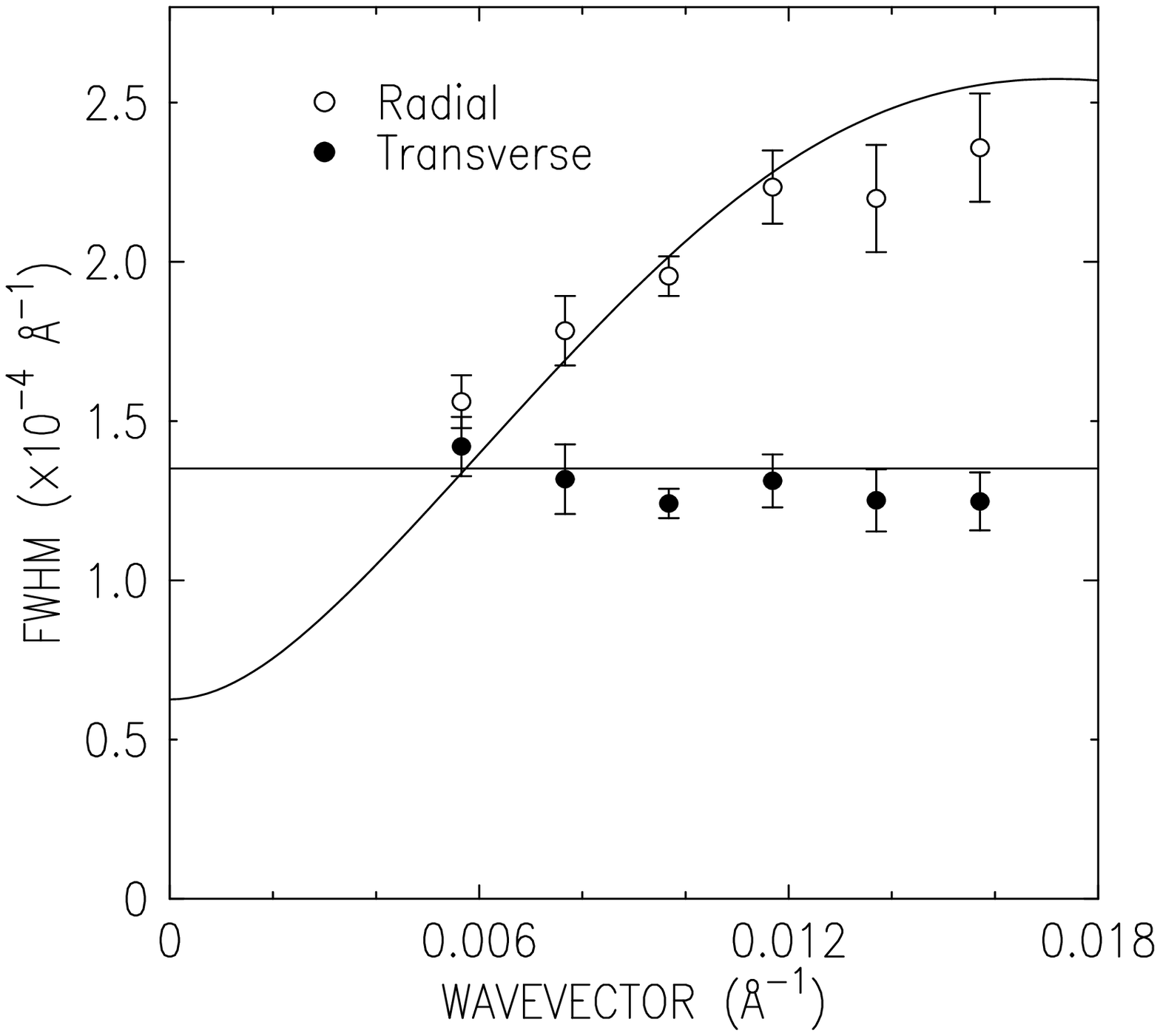,width=3.1in}
\end{center}
\figure{
FWHM of the background-subtracted autocorrelation
in the radial (open circles) and tangential (solid circles)
directions vs. wavevector.
\label{fig:FWHM}
}

\section{Discussion and conclusions}
\label{conclusion}
We now discuss what may be expected
for the FWHM of the background-subtracted
autocorrelation for our experimental configuration.
In our case, the tangential direction
is horizontal and the radial direction is vertical. Therefore,
the FWHM
of the background-subtracted
autocorrelation (in wavevector) in the tangential direction is
expected to be \cite{dainty:74}
\begin{equation}
\delta Q_{\theta} \simeq 5.57 / d_h,
\label{eq:tangential}
\end{equation}
where $d_h=4.2~\mu$m is the horizontal slit size.
Because the effective vertical size of the scattering volume increases 
with increasing scattering angle,
the speckle width in the vertical (radial) direction
tends to decrease with increasing wavevector \cite{chu:74}.
However, in the case that the bandwidth is non-zero,
the intensity in a given speckle,
which is actually at a fixed wavevector, becomes
distributed over a range of scattering angles corresponding to
the wavelength distribution. This tends to increase the speckle
width in the radial (vertical) direction.
On the basis of a Gaussian approximation for the scattering volume and
spectral distribution,
Abernathy \cite{abernathy:96} has derived that
the variation of the speckle radial width with wavevector is
given by
\begin{equation}
\delta Q_r \simeq
 5.57 \sqrt{ {
1+(0.064 d_v^2 Q^2+0.016W^2Q^4/k^2)( \delta \lambda / \lambda)^2 
} \over {
d_v^2 + W^2Q^2/k^2 + 0.016d_v^2W^2Q^4( \delta \lambda / \lambda )^2/ k^2
}},
\label{eq:radial2}
\end{equation}
where $d_v$ is the vertical slit size and $W$ is the sample thickness
(nominally 0.6~mm).
The FWHM predicted on the basis of Eq.~\ref{eq:tangential} and Eq.~
\ref{eq:radial2} are shown in Fig.~\ref{fig:FWHM} as solid lines.
The model provides a good description of the experimental
results.

Abernathy has also derived that the number of modes
increases with wavevector according to
\begin{equation}
M = M_0 
\sqrt{1+(0.064d_v^2Q^2 +0.016W^2Q^4/k^2) ( \Delta \lambda / \lambda )^2 }
\label{eq:MvsQ}
\end{equation}
This form with $M_0 = 2.3$ is shown as the solid line in Fig.\
\ref{fig:MvsQ}. Evidently,
Eq.\ \ref{eq:MvsQ} provides a fair description of the
wavevector dependence of the experimental data.
It is worthwhile to compare
the current result with earlier discussions 
\cite{dierker:95,albrecht:96,mochrie:96,abernathy:96,mainville:96},
which suggest that the maximum optical path
length difference should be less than the longitudinal coherence
length. This notion, in turn,
leads to the conditions that
$\lambda^2 / \Delta \lambda\ \gtrsim d_v Q / k$, and
$\lambda^2 / \Delta \lambda\ \gtrsim  W Q^2 / k^2$.
Formulated in the context of Eq.\ \ref{eq:MvsQ},
these conditions result in the criterion
$M \lesssim 2M_0$ under our experimental configuration, 
so that $\lambda^2/\Delta\lambda\gtrsim
\sqrt{2.52 d_v^2 Q^2 / k^2 + 0.63 W^2 Q^4 / k^4 }$.
As Fig.~\ref{fig:MvsQ} shows, this criterion is not satisfied in the range of
Q that we have studied, although the speckles are visible.
As for dynamical measurements, we were able to do XIFS studies
of colloidal suspensions for values of $M$ up to 11 using the 
current setup \cite{tsui:96}.
It follows that the conditions set based on considerations of 
optical path differences
are too stringent in accessing the feasibility of an XIFS experiment.

It remains to discuss what may be expected for $M_0$.
In considerations of the optical speckle produced by
perfectly-coherent laser illumination,
one is generally interested in the coherence area of the illuminated
sample volume at the detector, relative to the detector area.
Under these circumstances, the number of
modes, which contribute to the observed speckle statistics
is given by the ratio of the detector area to the 
coherence area of the illuminated sample volume at the detector.
In our case, the illuminated sample volume is determined by the
slits immediately upstream of the sample, and, as shown above,
the detector resolution is
smaller than the coherence area in question, so that the
detection scheme is essentially coherent.

However, our source is not perfectly coherent, and
to determine how many modes may be expected to
contribute to the observed intensity distribution,
as a result of finite source size,
we may compare the
coherence area of the source at the slit location to the actual slit area.
Under these circumstances, the number of
modes contributing to the observed speckle statistics
is given by the ratio of the slit area to the 
source coherence area at the slit location \cite{pusey:77}.
More precisely, we should compare the
source horizontal and vertical coherence lengths, respectively, 
to the horizontal and vertical sizes, respectively, of the slits.

To develop the appropriate quantitative comparision,
we employ several results given in Ref.~\onlinecite{dainty:74}.
We will initially
consider a one-dimensional source with
a Gaussian intensity distribution given by
\begin{equation}
I(s) = I(0) \exp( - \pi s^2 / \ell^2),
\label{eq:source}
\end{equation}
where $s$ is a coordinate in the plane of the source,
and $\ell$ specifies the source size.
Then, the number of modes transmitted through
a slit
of width $d$ at distance $R$ for x-rays of wavelength
$\lambda$ is
\begin{equation}
M_0 = 
(( \xi / d) {\rm erf}( \sqrt \pi d / \xi )
-( \xi^2 / \pi d^2)(1 - \exp( - \pi d^2 / \xi^2))^{-1}
\label{eq:integral}
\end{equation}
where
$\xi = \lambda R / \ell$ is the coherence length
\cite{dainty:74}.
Fig.\ \ref{fig:modelM} shows 
the number of modes according to
Eq.\ \ref{eq:integral} plotted versus $d/\xi$.
For $d / \xi \gtrsim 2$, the number of modes is
well approximated by $M_0 \simeq d/ \xi +0.37$.

In the case of the actual, two-dimensional source,
the number of contributing modes is the product
of two such terms, one for the horizontal direction ($M_h$) and one for the
vertical direction ($M_v$), {\em i.e.} $M_0 = M_v M_h$.
The nominal two-sigma vertical and horizontal source sizes at the NSLS are
17~$\mu$m and 820~$\mu$m, respectively, at the center of the wiggler.
However, because of the non-zero divergence of the electron
beam along the length of the wiggler, the effective two-sigma
vertical and horizontal source widths are estimated to be 36
and 1057~$\mu$m, respectively.
Assuming a Gaussian intensity distribution across the source,
it follows that the vertical and horizontal source sizes
are $\ell_v=45~\mu$m and $\ell_h=1325~\mu$m, respectively,
so that the vertical and horizontal coherence lengths are
$\xi_v = \lambda R / \ell_v = 99~\mu$m and
$\xi_h = \lambda R / \ell_h = 3.3~\mu$m, respectively.
For the vertical direction, $d_v / \xi_v \simeq 0.09$, and
we expect $M_v \simeq 1$;
for the horizontal direction $d_h / \xi_h \simeq 1.3$, and we
expect $M_h \simeq 1.7$.
Thus, we expect $M_0 \simeq 1.7$.
In fact, our data extrapolate to $M_0 \simeq 2.3$, which is only 35\%
larger than expected.

Finally, it is important to compare the observed rate of coherent 
x-ray photons to the value expected on the basis of the expected brilliance of
the X25 wiggler source, which is
$2 \times 10^{16}$~x-rays~per~0.1\%BW~${\rm s^{-1} mm^{-2} mrad^{-2}}$
at a ring current of 200~mA,
{\em i.e.} $3 \times 10^{9}$~x-rays~per~1.5\%BW~${\rm s^{-1} \AA^{-2} rad^{-2}}$.
The rate of coherent x-ray photons ($R_{coh}$) is the fraction of the
total rate ($R_{tot}$), that
is emitted into the coherence area of the source, {\em i.e.}
$R_{coh} = R_{tot} (\lambda / \ell_v \theta_v)  (\lambda / \ell_h \theta_h)
= \lambda^2 B = 8 \times 10^{9}$ coherent x-ray photons per second,
where $\theta_v$ and $\theta_h$ are the opening angles of
the wiggler in the vertical and horizontal directions, respectively,
and $B$ is the brilliance.
We should not expect to realize this rate in the present experiment,
in part,
because the vertical
coherence length ($99~{\rm \mu m}$) is larger by a factor of 11
than the vertical slit size ($8.9~{\rm \mu m}$).
In addition, because of filters, beryllium windows, and air in the x-ray path,
there may be a factor of 3 (at most) attenuation along the beamline,
and the reflectivity of the multilayer pair is 50\%.
Thus, we should ideally expect $1 \times 10^8$ coherent photons per second.
This may be compared with $4 \times 10^7$ photons passing our
slits.

\begin{center}
\epsfig{figure=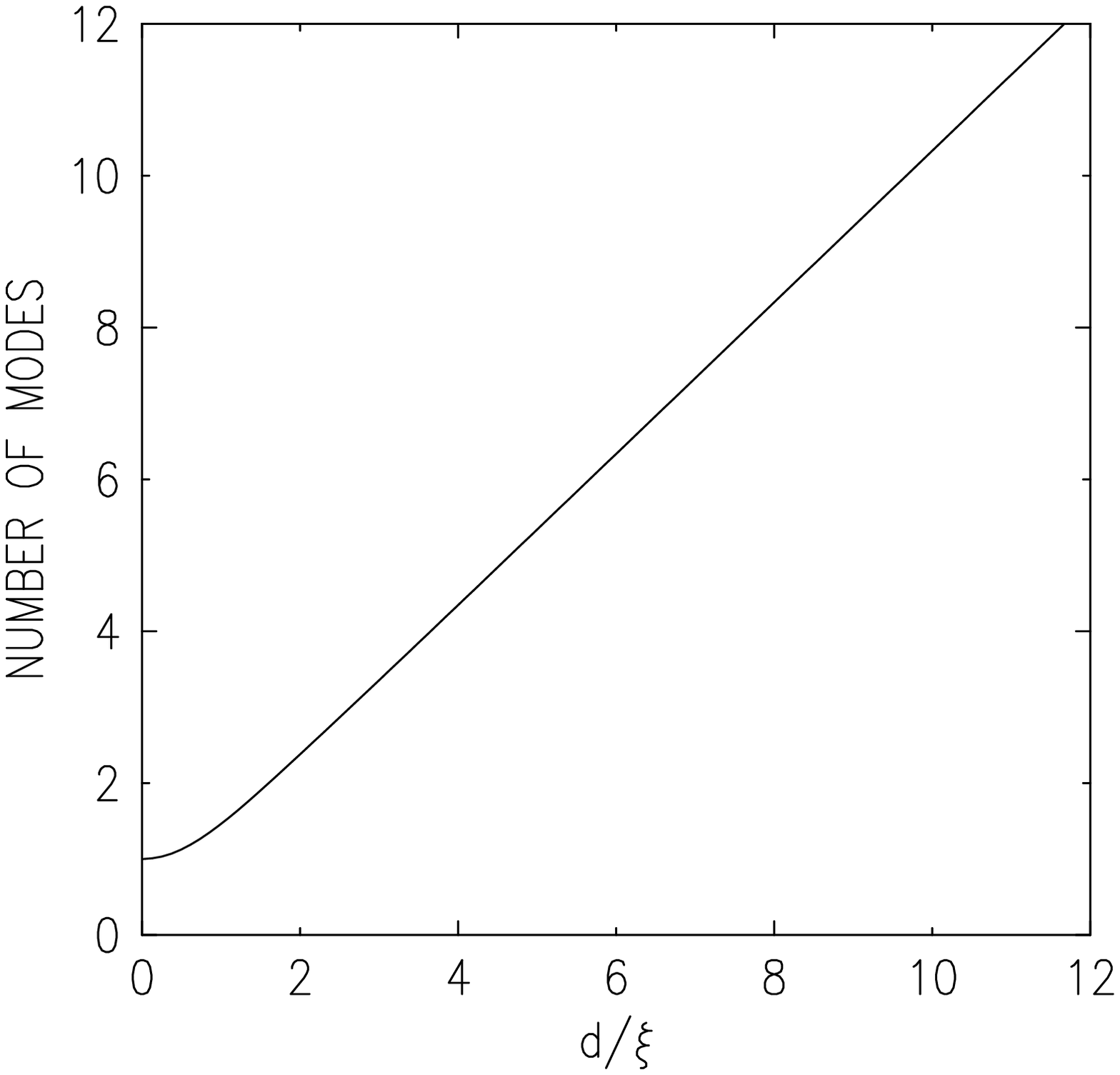,width=3.1in}
\end{center}
\figure{
Number of modes vs. $d/ \xi$,
according to Eq.\ \ref{eq:integral}. 
\label{fig:modelM}
}

In conclusion, we have performed a statistical analysis of
speckle from an aerogel sample produced at beamline X25 at the
NSLS. The experimental scheme described in here provides a 
convenient means to monitor the coherence of an x-ray beam. 
The results indicate that the coherence is within 35\%
of what is expected, and that the number of coherent
photons is a factor of $\sim 2$ smaller than expected
on the basis of the expected brilliance.
Finally, a number of formula have been collected that may be helpful in
considerations of the feasibility and optimization
of XIFS studies using large bandwidth radiation.

We would like to thank D. Abernathy, S. Brauer, S.B. Dierker, G. Gr{\"u}bel,
S. Hulbert,
F. Livet, L. Lurio, J. Mainville, A.M. Mayes, I. McNulty, N. Mulders,
A. Sandy, G.B. Stephenson, M. Sutton, G. Swislow, and M. Yoon
for fruitful interactions.
Work at MIT was supported by the NSF (Grant DMR-9312543).
Work at the NSLS is supported by the U.S. DOE under
Contract No. DE-AC0276CH00016.


\bibliographystyle{simon}
\bibliography{spec}

\end{document}